# Distance measurements between carbon and bromine using a split-pulse PM-RESPDOR solid-state NMR experiment

M. Makrinich[1], M. Sambol[2], A. Goldbourt[1*]

[1]*School of Chemistry, Tel Aviv University, Tel Aviv, 69978, Israel*
[2]*Ami"t Science High school, Kfar Batia, Ra'anana, Israel*
[*]*amirgo@tauex.tau.ac.il*

## Abstract

Solid-state Nuclear Magnetic Resonance have been long used to probe atomic distances between nearby nuclear spins by virtue of the dipolar interaction. New technological advances have recently enabled simultaneous tuning of the radio-frequency resonance circuits to nuclei with close Larmor frequencies, bringing a great promise, among other experiments, also to distance measurements between such nuclei, in particular for nuclei with a spin larger than one-half. However, this new possibility has also required modifications of those experiments since the two nuclei cannot be irradiated simultaneously. When measuring distances between a spin S=1/2 and a quadrupolar spin (S > ½), this drawback can be overcome by splitting the continuous-wave recoupling pulse applied to the quadrupolar nucleus. We show here that a similar adjustment to a highly-efficient phase-modulated (PM) recoupling pulse enables distance measurements between nuclei with close Larmor frequencies, where the coupled spin experiences a very large coupling. Such an experiment, split phase-modulated RESPDOR, is demonstrated on a $^{13}C$-$^{81}Br$ system, where the difference in Larmor frequencies is only 7%, or 11.2 MHz on a 14.1T magnet. The inter-nuclear distances are extracted using an unscaled analytical formula. Since bromine usually experiences particularly high quadrupolar couplings, as in the current case, we suggest that the split-PM-RESPDOR experiment can be highly beneficial for the research of bromo-compounds, including many pharmaceuticals, where carbon-bromine bonds are prevalent, and organo-catalysts utilizing the high reactivity of bromides. We show that for Butyl triphenylphosphonium bromide, solid-state NMR distances are in agreement with a low-hydration compound rather than a water-caged semi-clathrate form. The split-PM-RESPDOR experiment is suitable for distance measurement between any quadrupolar↔spin-1/2 pair, in particular when the quadrupolar spin experiences a significantly large coupling.

Keywords: Solid-state NMR; dipolar interaction; quadrupolar interaction; phase-modulated RESPDOR; distance measurement; tetra-n-butylammonium bromide hydrates; Br-81

## 1. Introduction

The introduction of the Rotational Echo Double Resonance (REDOR) [1] experiment in 1989 by Gullion and Schaefer has opened the path to accurate and robust distance measurements by magic-angle spinning solid-state NMR (MAS ssNMR) techniques, and these have proven extremely useful in numerous studies since. In the REDOR experiment, pairs of rotor-synchronized π pulses are applied every rotor period thereby decoupling all but the dipolar interaction ('S' signal). When the pulses are omitted from the coupled spin, a reference signal is obtained ("$S_0$" signal). Propagation of time is achieved by incrementing the number of these pairs of π pulses, and therefore the increase of the REDOR fraction 1-$S/S_0$ is a measure of the dipolar interaction, that is, the inter-nuclear distance. Different versions of the experiment located the pulses either one per rotor period on every channel, or all but a single refocusing pulse at one of the channels. Recent developments also facilitated efficient applications of the experiment at spinning speeds that exceed the available practical nutation frequencies[2].

Despite the great robustness of the experiment, the requirement for the pairs of π pulses to fully invert the spin coherence in order for the interactions to fully be decoupled has prevented these kinds of measurements to be conducted on nuclei possessing large anisotropic couplings, such as the quadrupolar interaction, chemical shift anisotropy, and paramagnetically broadened species. This obstacle has led to further developments of REDOR-based experiments that could be (with varied efficiency) applied to quadrupolar nuclei as well. REAPDOR [3], LA-REDOR [4] and S- or R-RESPDOR [5,6] are some of the experiments that were developed and allowed the expansion of distance measurement potential to a larger variety of pairs of nuclei. All these experiments utilize a continuous-wave (cw) single pulse on the coupled quadrupolar nucleus lasting between a third and two rotor periods, and a spin-1/2 recoupling period where either pairs of π pulses or symmetry-based recoupling is used. Nevertheless, for nuclei possessing quadrupolar interactions exceeding several megahertz, these widely and successfully used approaches begin to deteriorate in their efficiency and accuracy and hence many spin systems remained challenging.

A few examples for nuclei that tend to experience large quadrupolar interactions ($\nu_Q$ greater than ~5 MHz) are halogens[7,8], and various metals, all appearing in many biological, catalytic, and other important systems. Clearly there was a benefit to be gained from developing an experiment that could recouple distances to quadrupolar nuclei with larger quadrupolar frequencies more efficiently. Such improvement was achieved with the replacement of the cw pulse with a phase-modulated pulse, an experiment termed initially PM LA-REDOR, and later in accordance with the general acronym approach, PM-RESPDOR [9]. Indeed, we have recently shown that the PM-RESPDOR experiment enabled us to retrieve distances in two challenging systems - a $^{13}$C-$^{209}$Bi (spin-9/2, $\nu_Q$=10.7 MHz) distance was obtained in bismuth-acetate, and a $^{31}$P-$^{79/81}$Br (spin-3/2, $\nu_Q$= 4.7 MHz for $^{81}$Br and 5.6 MHz for $^{79}$Br) distance in a different compound[10]. Combining the PM pulse with the symmetry-based recoupling for the spin-1/2 nucleus, we have recently demonstrated the advantages of this pulse in the measurements of $^{13}$C-$^{14}$N distances as well, in an experiment acronymed PM-S-RESPDOR[11].

The main limitation in the distance measurement experiment is the inability of standard probes to tune to two channels simultaneously if the difference between their Larmor frequencies is in the order of 1-25 MHz. A notable innovation of this kind was introduced by Haase, Curro and Slichter in 1998, when they suggested a new probe design that allowed measuring such spin pairs[12]. In 2002, $^{13}$C-$^{27}$Al REAPDOR distance measurements were demonstrated by Wullen and co-workers using a homebuilt double-resonance set-up. Those distances were used to characterize the geometry of methanol in HZSM-5 zeolite in order to explore methanol to gasoline reaction pathways[13]. A different combination of frequencies, $^7$Li-$^{31}$P, was utilized by van Wullen et al. to study lithium-phosphate glasses[14].

A commercial frequency splitter is now available for splitting a single frequency into a double-resonance circuit, thus allowing a broader usage of distance measurements between nuclei with close Larmor frequencies. However, when using this new device, REDOR-based approaches involving simultaneous irradiation on both channels still requires an adjustment, as it only enables irradiation of one of the channels at a time. Such an adjustment was proposed for the S-RESPDOR experiment in 2012,[15] where $^{13}$C-$^{27}$Al distances were measured by splitting the saturation pulse applied on $^{27}$Al into two equal parts, and fitting the π pulse on $^{13}$C in the gap. This work paved the way for S-RESPDOR distance measurement experiments to be used for systems with close Larmor frequencies, such as the first $^{51}$V-$^{13}$C distance measurement by NMR[16]. Additional studies demonstrated correlation measurements between nuclei with close

Larmor frequencies such as DNP enhanced $^{27}$Al–$^{13}$C 2D D-HMQC that was used for probing the structure of a metal-organic framework material[17], and $^{27}$Al–$^{13}$C HETCOR D-HMQC and J-HMQC 2D experiments conducted on two co-catalysts that have importance for the polymerization process of olefins[18]. Recently we demonstrated how $^7$Li-$^{31}$P 2D TEDOR and REDOR experiments can be utilized to define the binding environment of the mood stabilizer Li in ATP[19].

In all the studies mentioned above the quadrupolar spins had relatively small couplings (~0.4-2 MHz for Al, 0.3 MHz for $^{51}$V, 25 kHz for $^7$Li), allowing measurements of inter-nuclear distances, and acquisition of correlation experiments by techniques involving direct inversion pulses or using cw saturation pulses. $^{81}$Br is an example of a spin that usually possesses much stronger quadrupolar couplings, and its abundance in pharmaceuticals, in organo-bromine compounds, and in bromide-based catalysts makes the ability to measure $^{13}$C-$^{81}$Br distances highly desirable. Since $^{13}$C and $^{81}$Br have only a difference of 11.2 MHz (6.9%) in their Larmor frequencies (on a 14.1 T magnet - 600 MHz proton frequency), carbon-bromine distances have not been reported up to date by NMR techniques. In this work we show that a split PM-RESPDOR experiment can be used with a frequency splitter to successfully recouple $^{13}$C-$^{81}$Br distances, opening its usage potential to nuclei with close Larmor frequencies. We show that this approach is sufficiently robust so that the distance can be obtained with a universal analytical formula fitted with a single parameters, the distance.

## 2. Methods

### 2.1. Materials:

Butyl triphenylphosphonium bromide ($C_{22}H_{24}BrP$, BrBuPPh$_3$) and tetra-n-butylammonium bromide ($C_{16}H_{36}BrN$, TBAB) were purchased from Acros Organics and Chem-impex intl respectively, and used without further modifications.

### 2.2. NMR experiments:

All experiments were performed on a Bruker AvIII spectrometer with a magnetic field of 14.1 T yielding Larmor frequencies of 599.9 MHz for $^1$H, 242.8 MHz for $^{31}$P, 162.0 MHz for $^{81}$Br, and 150.8 MHz for $^{13}$C. Triple resonance mode was used for all $^{31}$P-$^{81}$Br distances. The $^{79}$Br nuclear quadrupolar coupling constant of both compounds was determined by employing WURST-QCPMG[20] on a static sample. The value of $C_Q$=11.3 MHz for BrBuPPh$_3$ is in agreement with literature[8]. For TBAB we determined a value of 12.6 MHz, corresponding to $C_Q$=10.55 MHz for $^{81}$Br. Spectra and fits are shown in Figure S1 of the supporting information (SI). $^{13}$C-$^{81}$Br distances in TBAB were measured using the split PM-RESPDOR experiment with the probe set to a double resonance ($^1$H-X) mode with a frequency splitter obtained from NMR Service, which allowed measuring both $^{13}$C and $^{81}$Br on the same channel (X). $^{31}$P-$^{81}$Br distances were measured in BrBuPPh$_3$ both with PM-RESPDOR and the split PM-RESPDOR methods omitting the frequency splitter, for comparison. All PM-RESPDOR experiments were performed using only two π-pulses on the spin-half nucleus during the PM saturation interval [10]. The pulse program suitable for the topspin v3 software available in Bruker spectrometer appears in the SI.

Experimental parameters are described in the table below:

| Compound and experiment \ Parameters | BrBuPPh3, $^{31}$P{$^{81}$Br} PM-RESPDOR | BrBuPPh3 $^{31}$P{$^{81}$Br} split PM RESPDOR* | TBAB $^{13}$C{$^{81}$Br} split PM-RESPDOR** |
|---|---|---|---|
| spinning speed | 11 kHz | 11 kHz | 11 kHz |
| "Overall" PM pulse length | 10 Tr | 10 Tr*** | 10 Tr*** |
| Actual PM pulse length | 10 Tr | 9.5834 Tr, 9.5867 Tr | 9.5526 Tr |
| recovery time (~5$T_1$) | 70 s | 70 s | 35 s |
| number of scans | 8 | 8, 32 | 128 |
| acquisition time | 30 ms | 30 ms | 38 ms |
| $^1$H π/2 pulse power**** | 100 kHz | 100 kHz | 100 kHz |
| $^1$H CP power | 61 kHz | 61 kHz, 63 kHz | 53 kHz |
| Spin 1/2 ($^{13}$C for TBAB, $^{31}$P for BrBuPPh3) CP power | 47 kHz | 47 kHz | 37 kHz |
| CP duration | 3 ms | 3 ms | 2 ms |
| SWF-TPPM $^{21}$ decoupling power | 100 kHz | 100 kHz | 100 kHz |
| Spin 1/2 ($^{13}$C for TBAB, $^{31}$P for BrBuPPh3) π pulses power | 47 kHz | 47 kHz, 48 kHz | 37 kHz |
| Split length | --- | 26.3 μs | 26.3 μs |
| PM pulses power (on $^{81}$Br)***** | 37 kHz | 37 kHz | 41 kHz |
| Phase Cycling on the spin-half π pulses ($\phi_3$ in fig. 1) $^{22}$ | XY64 | XY64 | XY8 |

**Table 1:** Experimental parameters for butyltriphenylphosphonium bromide (BrBuPPh3) and Tetra-n-butylammonium bromide (TBAB) PM-REPSODOR experiments.

*Some rows have two values, because points of longer dephasing times on the recoupling curve were measured few days later, after a new optimization on the power levels of $^{31}$P and $^1$H. The number of scans was also increased from 8 to 32.
**With a frequency splitter
***Including the splitting. In other words, this is the time duration between the red dots in fig. 1.
****Power levels are given by $\nu_1$.
****Power level for $^{81}$Br were estimated from KBr (KBr was also used to determine the $^{81}$Br carrier frequency). The nutation frequency is twice that value, $\nu_{nut}=(S+1)\nu_1$.

### 2.3. Simulations:

Numerical simulations of $^{13}$C{$^{81}$B} and $^{13}$C{$^{27}$Al} PM-RESPDOR and its split version were performed using SIMPSON$^{23}$ version 4.1.2 on a 16-core Ubuntu-linux system. Ideal π pulses were used on $^{13}$C, except the two pulses between the red dots in Figure 1 (the echo pulses). A quadrupolar coupling constant ($C_Q$) of 9 MHz was used for $^{81}$Br. The $^{13}$C-$^{81}$B simulations were performed with a dipolar coupling constant of 132 Hz corresponding to a distance of 3.95 Å, and with a spinning speed ($\nu_R=1/T_R$) of 11 kHz, using 18 Euler γ-angles and 320 α/β angles employing the 'Repulsion' method for powder averaging. A radio-frequency (rf) irradiation power corresponding to a nutation frequency of 40 kHz was used for all the pulses on both $^{13}$C and $^{81}$B, leading to a π pulse length of 12.5 μs for $^{13}$C. Following the actual experiments, a splitting gap of 0.3$T_R$-1μs was used, leading to a rescaling the PM pulse segments by 0.95625 and to an actual PM pulse length of 9.5625 $T_R$. Three-spin simulations employing two $^{81}$Br spins, shown in Figure S2, were performed using $C_Q$=11 MHz and with two different dipolar coupling constants, one constant at 130 Hz and the second varying between 130 and 24 Hz.

$^{13}$C{$^{27}$Al} split-PM-RESPDOR simulations, representing a spin-1/2 ↔ spin-5/2 spin pair, were performed with a varying dipolar interaction constant, quadrupolar coupling constant, and splitting gap, as described in the inset of figure 2. All other parameters were the same as for the $^{13}$C{$^{81}$B} simulations.

# 3. Results and discussion

## 3.1. The split PM-RESPDOR experiment

The split PM-RESPDOR experiment shown in Figure 1 resembles the original PM-RESPDOR experiment[9], with the exception of the phase-modulation period, marked by the red dots. While in the original pulse sequence the phase-modulated saturation pulse lasts for the whole "confined by the red dots" period, here it only starts after the spin-half π pulse has ended and lasts a period of $PM1 = \frac{1}{2}(n \cdot 10T_R - \tau_\pi - 0.3T_R)$ where $\tau_\pi$ is the π pulse duration on the spin-half nucleus, $0.3T_R$ is the time interval chosen for the splitting (other time periods might have worked as well, see below), and $n$ is the PM saturation pulse length in units of $10T_R$ including the splitting (in other words, $n \cdot 10T_R$ is the interval between the red dots in fig. 1). In all our experiments here $n=1$ was sufficient however when $C_Q$ is larger, $n > 1$ may be required, or in general $mT_R$ rotor periods can be utilized with $m > 10$. After the $0.3T_R$ gap another PM1 saturation pulse is applied, and ends exactly where the second red dot is located. Both PM1 blocks are of the same length but are continuous in the order of phases; they consist of two kinds of pulses - one with length of $\frac{2*PM1}{10T_R} \cdot 0.75T_R$ and a constant phase of 225°, and the other one that lasts $\frac{2*PM1}{10Tr} \cdot 0.109375T_R$, with phases that vary between 2.2°-348.1°, as described in the PM-RESPDOR original work[9]. Here the duration of these two pulses was slightly rescaled by $\frac{2*PM1}{n*10Tr}$, in order to make room for the splitting gap. In addition, $\frac{2*PM1}{n*10Tr}$ was multiplied by $n$ in order to extend the time interval between the red dots to last $n \cdot 10T_R$ instead of $10T_R$, similarly to the extension that was used in previous work[10]. In our case $n=1$.

In this version of PM-RESPDOR, only two π pulses are applied on the spin-half nucleus during the quadrupolar saturation pulse, which makes interlacing them in the non-irradiative intervals easier[10]. The first π pulse is applied at the beginning of the "overall" saturation period (marked by the first red dot), and the second one starts exactly $5n$ rotor-periods afterwards (as in the original sequence).

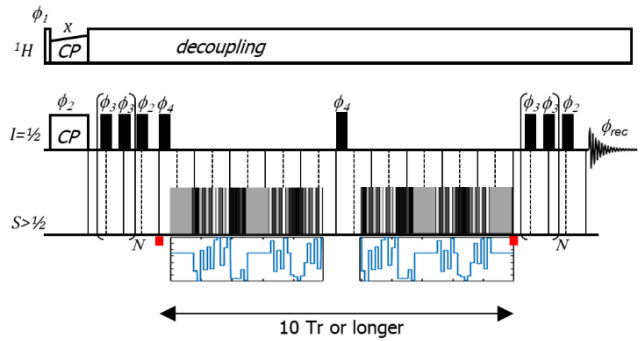

**Figure 1.** Split PM-RESPDOR pulse sequence. Filled bars on the half-spin are π pulses. The quadrupolar S-spin pulse is color-coded in a gray scale according to phase, which appears in the plot below. The actual phase values can be found in the SI and elsewhere[9] and their lengths are described above. The length of $10T_R$ may be extended to more rotor periods (either n·$10T_R$ with integer $n$, or any multiple of the rotor period with proper adjustment of the PM pulse). Between the red dots it is safe to assume that coherent dipolar evolution is not taking place.

## 3.2. Simulating split PM-REPSDOR – the effect of the gap

In order to examine the effect of introducing a gap into the pulse, we simulated several cases including variations in the gap length, quadrupolar coupling strength, and the dipolar coupling strength. All cases, summarized in **Figure 2**, show that the effect of the gap is minimal and that the experiment can still safely be fit with the analytical Bessel function with a single parameter fit (be it the distance or the dipolar interaction).

In Figure 2a, we show the entire PM-RESPDOR curve for a spin-3/2 experiencing a quadrupolar frequency $\nu_Q = C_Q/2 = 4.5$ MHz. It is practically impossible to decipher the PM-RESPDOR curve from the split version and from the analytical Bessel function given by equation (1)[24].

$$\frac{\Delta S}{S_0} = na \cdot \frac{1}{4}\left\{3 - \frac{\pi\sqrt{2}}{16}\sum_{k=1}^{3}[6 - 2(k-1)]J_{\frac{1}{4}}(k\sqrt{2}(2N+2)DT_R)J_{-\frac{1}{4}}(k\sqrt{2}(2N+2)DT_R)\right\} \quad (1)$$

Here '$na$' is the natural abundance of the quadrupolar-spin isotope, and $(2N+2)T_R$ is the total dephasing time and does not include the PM pulse time itself.

In Fig. 2b we varied the quadrupolar coupling strength experienced by a spin-3/2 and by a spin-5/2. For a spin-5/2 we fit the data with the Bessel function adjusted for this system (eq. 2):

$$\frac{\Delta S}{S_0} = na \cdot \frac{1}{6}\left\{5 - \frac{\pi\sqrt{2}}{24}\sum_{k=1}^{5}[10 - 2(k-1)]J_{\frac{1}{4}}(k\sqrt{2}(2N+2)DT_R)J_{-\frac{1}{4}}(k\sqrt{2}(2N+2)DT_R)\right\} \quad (2)$$

The fit values, denote by $D_{fit}$, are compared to the actual dipolar coupling we simulated (D=130 Hz), and this fit is repeated for both split and non-split versions. In all cases, the two methods are hardly separable, and all fit well to D, never below 93% of the expected value, and inline with results obtained for the non-split PM-RESPDOR technique. As shown before[10], when the quadrupolar coupling strength becomes large, the PM pulse needs to be extended in order to obtain an efficient curve that can be fit with Eqs. 1 and 2. This is demonstrated for a spin-5/2 with $\nu_Q = 3C_Q/20 = 10.2$ MHz where a short PM pulse of $10T_R$ does not provide a good distance estimate, but with $30T_R$ the value is again within 10% of the actual dipolar coupling constant (more accurately, 4%), and is accurate also for the split version. Similar simulations for a spin-3/2 show that a fit to Eq. 1 can be obtained even under very strong couplings provided that the PM pulse is extended. Even an attempt to simulate a recoupling curve with $C_Q$=200 MHz results in significant dephasing, which improves with the extension of the PM pulse (see also Figure S3 in the SI). With such strong couplings other limitations can appear such as very strong CSA-quadrupolar broadening[25] and strong signal dephasing due to relaxation.

Figure 2c shows that even if we increase the gap up to three rotor periods, the experiment is still viable and the error in the fit is still smaller than 10%. Up to a single rotor period, the accuracy is even higher, below 4%. This result is important since it allows one to perform a spin-1/2 selective experiment by incorporating a long weak selective pulse on this spin, without compromising the efficiency of the experiment. Finally, a similar fit procedure was repeated for variations of the dipolar coupling value yielding again negligible dependence on the small gap of 26.3μs.

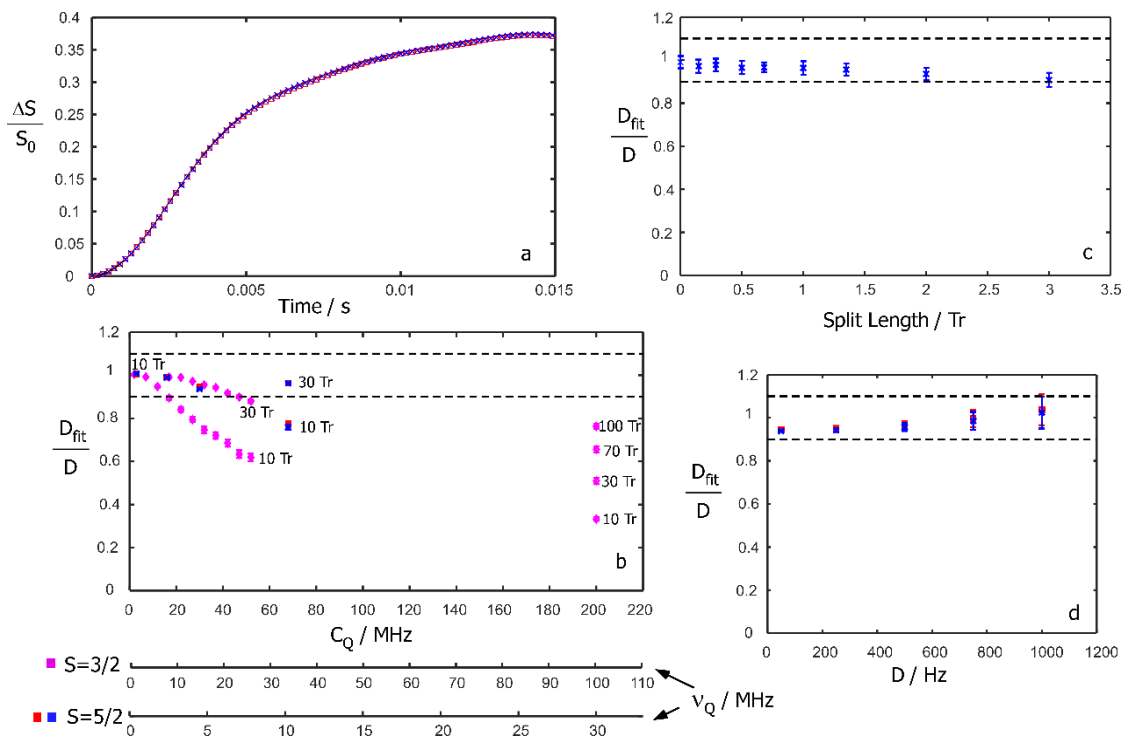

**Figure 2.** Simulations of the split-PM RESPDOR experiment. (a) A recoupling curve obtained for a spin-3/2 with (blue cross) and without (red squares) the gap, and the corresponding Bessel function (Eq. 1, solid black line) with D=132 Hz, $\nu_Q$=4.5 MHz, a spinning speed of $T_R^{-1}$=11 kHz, and a gap of 26.3µs (0.3$T_R$-1µs). (b) Simulations of the recoupling curve for a spin-1/2 coupled to a quadrupolar spin-5/2, with (blue cross) and without (red square) the gap, were fit using the Bessel function in Eq. 2 (*na*=1) and the values of $D_{fit}/D$ are plotted for different quadrupolar coupling constants. The simulations were done with D=130 Hz and fitting was performed for simulation points at the entire rise of the curve up to t=11.3 ms. The gap is 26.3µs. For the strongest coupling for a spin-5/2 the PM pulse was also extended to 30$T_R$ (marked) since a short 10$T_R$ pulse is insufficient to saturate the entire quadrupolar lineshape and thus does not provide a reasonable fit. Simulations and fits to Eq. 1 to a coupled spin-3/2 are also shown on the same plot for an extensive range of $C_Q$ values up to 50 MHz using PM pulse lengths of 10$T_R$ and 30$T_R$, and include an additional point at CQ=200 MHz for a range (10-100$T_R$) of PM pulse lengths (see also Fig. S3). (c) Simulations of the spin-5/2 system in (b) as a function of the gap with D=750Hz, $\nu_Q$=4.5 MHz. (d) Simulations of the spin-5/2 system in (b) as a function of the dipolar coupling constant with $\nu_Q$=4.5 MHz, and a gap of 26.3µs.

### 3.3. Experimental verification – $^{31}$P-$^{81}$Br distance measurements.

Initially we tested the performance of the split-PM pulse with respect to a known system, which has been measured already before. We used a regular commercial triple-resonance 4 mm probe to measure a $^{31}$P-$^{81}$Br distance in BrBuPPh$_3$. **Figure 3** shows the $^{31}$P{$^{81}$Br} split-PM- and regular PM-RESPDOR experimental ∆S/S$_0$ fraction curves and the corresponding single spin-pair distance fits. For a spin-3/2, the experimental data points were fit to the analytical Bessel function given by equation (1). Here we used *na*=0.495 to account for the natural abundance of bromine-81. Both curves yield almost an identical distance (within the error range), 4.57(±0.05)Å for the regular experiment without splitting the pulse, and 4.61(±0.02)Å for the split-PM experiment. Both match excellently to the crystallographic distance of 4.54Å[26]. This comparison validates experimentally the use of the pulse splitting to two parts and we could thus safely proceed to measure $^{13}$C-$^{81}$Br distances.

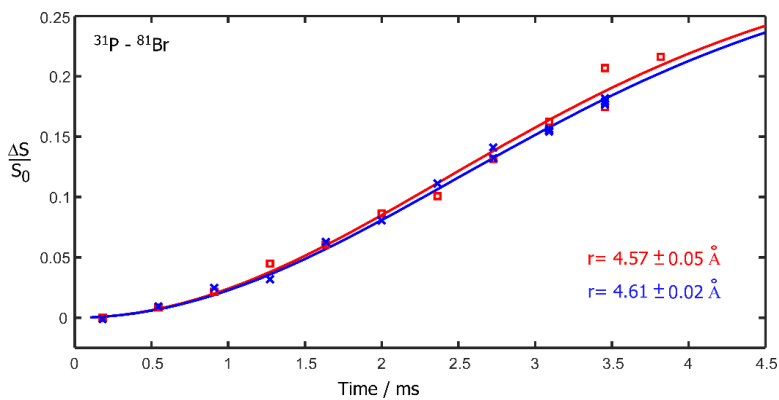

**Figure 3.** $^{31}P\{^{81}Br\}$ PM-RESPDOR (red squares) and split-PM-RESPDOR (blue crosses) fraction curves of BrBuPPh$_3$. The experimental points were fitted to equation (1) using 'cftool' of MATLAB.

### 3.4. Close frequencies: $^{13}C$-$^{81}Br$ distance measurements and identification of the hydrate form of tetra-n-butylammonium bromide

Measuring distances between carbon and bromine is usually hampered by the fact that their Larmor frequencies are either too close $|\nu_0^{13C} - \nu_0^{79Br}| = 0.547\ MHz$ (at 14.1T) so that off-resonance effects may impact the experiment, or close enough $|\nu_0^{13C} - \nu_0^{81Br}| = 11.247\ MHz$ to preclude triple-tuning on a conventional probe. Another limitation is on the efficiency of the experiment imposed by the large quadrupolar coupling constant of bromine, usually in the order of several Megahertz. Thus, in order to probe carbon-bromine distances, we combine the frequency splitter with the split-PM-RESPDOR technique.

We applied the experiment to the quaternary ammonium salt tetra-n-butylammonium bromide (TBAB), **Figure 4**. The positive charge on the nitrogen in this material is compensated for by the negatively charged bromine ion however, the position of this bromine ion depends strongly on the hydration state of this highly hygroscopic, potentially semi-clathrate hydrate crystal. Hydrated forms are known that include between $2\frac{1}{3} - 38$ water molecules[27,28] and the bromine ion position varies. Also, those structures show some disorder in the position of the carbon chains, and they can take two alternate conformations in some of the hydrates.

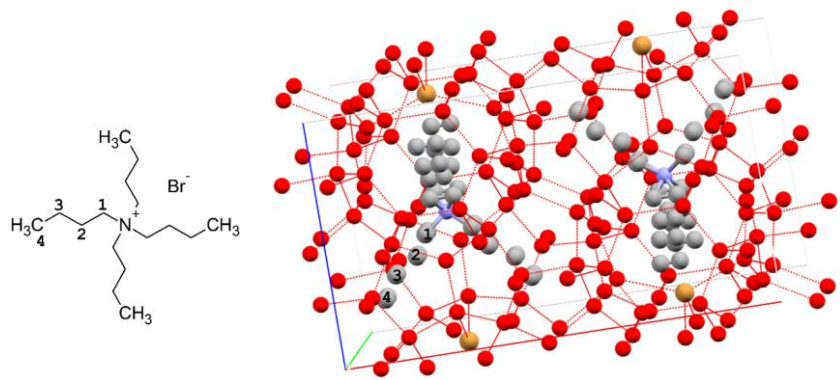

**Figure 4.** The chemical structure (left, generated by ChemSketch) and a single unit cell (right, generated with Mercury version 3.10.3) of TBAB•38H$_2$O[29]. In one unit cell, two TBAB molecules are caged by the water molecules. Carbon numbering (shown on plot) is from the nitrogen outwards making the methyl group C4. The two possible positions of the chains are manifested by the doubling of the carbons in the unit cell. The Br-C distances increase from the C4 methyl group (4.04Å) to C1 (5.01Å) and are shorter for the bromines on the top left and down-right of the image. In the plot, oxygen in red, carbon in gray, nitrogen in cyan, bromine in orange.

The $^{13}$C spectrum of TBAB shown in **Figure 5** contains four main shifts, corresponding to the different carbons on the butyl chain. Some of the signals further split showing some fine structure. The fine structure may be attributed to the disorder of the chain, or to the same carbons at different chains, however since we cannot explicitly assign them, and since those split signals show almost identical recoupling curves (3.86±0.06Å, 3.83±0.04Å for C2, 3.9±0.1Å, 4.03±0.09Å, 3.95±0.08Å for C3, 4.21±0.06Å, 4.34±0.07Å, 4.16±0.06Å for C4, see Figure S4 in the SI) we analyzed recoupling curves from broadened spectra (broadening of 100 Hz) where the signals merge representing an average over the different chains..

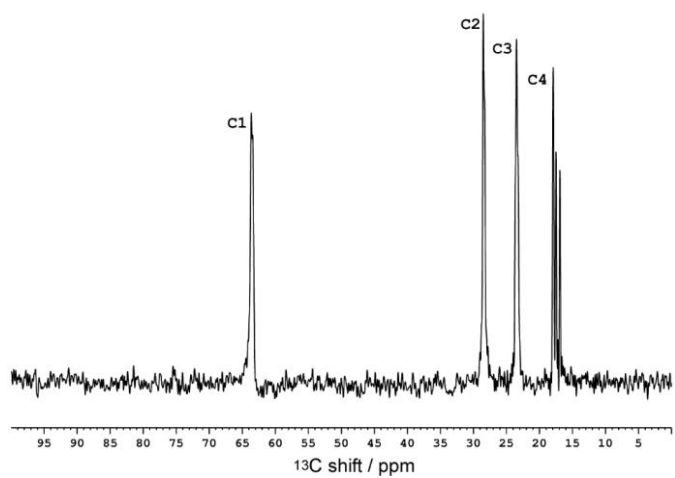

**Figure 5.** {$^1$H}$^{13}$C-CPMAS spectrum of TBAB showing the four carbons on the chain. The spectra were slightly line-broadened with 20Hz. The peaks correspond from left to right to C1 (63.6 ppm), C2 (28.4 ppm), C3 (23.4 ppm), C4 (17.9, 17.4, 16.8 ppm). Labeling is as shown in Figure 4. Chemical shifts are reported with respect to Adamantane methylene line at 40.48 ppm.

Split-PM REPSDOR $^{13}$C{$^{81}$Br} recoupling curves and fits to equation (1) with '*na*'=0.495 are shown in **Figure 6**. The data (up to 4.5 ms) fit well to single $^{13}$C-$^{81}$Br spin pairs and the distances are for C1 3.98(±0.04)Å, for C2 3.84(±0.04)Å, for C3 3.95(±0.04)Å and for C4 4.21(±0.03)Å. Both the order of the distances as well as their absolute values do not fit the semi-clathrate structure shown in **Figure 4**. They also do not fit the hydrate with 32.5 water molecules where the closest distance for C1 is 6.6 Å, and not the one with 24 water molecules, where very short C-Br distances are encountered (2.8Å, 3.1Å).[28] The distances fit well to the lowest hydrate containing 2 1/3 water molecules[27]. Interestingly, in this compound the water and bromine ions are caged between several TBAB units, unlike the structure shown above for TBAB•38H$_2$O.

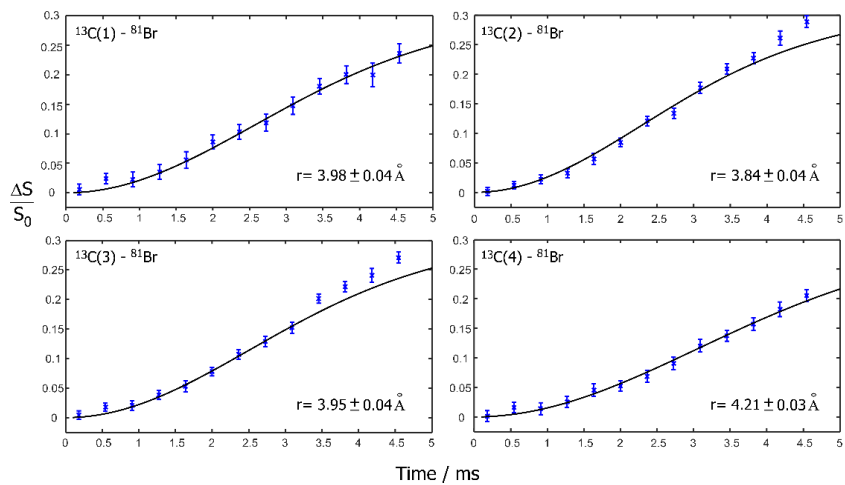

**Figure 6.** Split-PM RESPDOR $^{13}$C{$^{81}$Br} recoupling curves of the compound TBAB, and fits to equation (1) are shown for the four carbons in the chain. Distances and error estimations were obtained by the 'cftool' for non-linear fit in MATLAB.

Analysis of C-Br distances below 4.5Å in TBAB•2$^1/_3$H$_2$O show that C1 is at distances of 3.9-4.4 Å, C2 at 3.7-4.2 Å, C3 is at 3.9-4.3 Å, and C4 is at distances of 3.9-4.5 Å. Some of these distances are reported to a disordered bromide site (Br2) and thus produce some distance distributions. In other cases a carbon may be in proximity to two bromine ions but such proximity results in an error of no larger than 15% in the distance when the two distances are equal, and even a difference as small as 0.5 Å reduces this error to below 10%.[30] We further examined these errors for a spin system containing two bromines with natural abundance distribution, as shown in the SI, figure S2. When fitting the three-spin simulations with the analytical Bessel function (Eq. 1), the error in distance is not higher than ~0.2-0.3Å for the TBAB system. For example, the curve for C1 in Figure 6 that was fit with a single Br at a distance of ~4.0Å, could be fit with two bromines at 4.1Å and 5.1Å, keeping the result for the closest bromine sufficiently accurate.

In the crystal structure we can find C1 at a distance of 3.9Å, and then C1 on a different chain having two distances of 4.1Å and 4.3Å. Summing on these contributions, and considering the error, we still remain with results that fit our curves. C3 also exhibits two close distances at 4.6Å to Br1 and between 3.8-4.3Å to the disordered bromine resulting again in a small potential under-estimation of the NMR distance but still in agreement with our data. Thus, although the curves corresponding to our sample, used 'as is', represent a distribution of distances around some values, with combinations of one or two nearest bromides, the results are very close to TBAB•2 1/3H$_2$O, with C2 indeed showing the shortest distance. In no case do we observe a distance of >5Å or <3 Å as in the semi-clathrate materials with 20 water molecules and more, and we can safely state that our sample corresponds to the reverse structure of TBAB•2 1/3H$_2$O showing the octahedral arrangement of the six organic moieties, as shown in Figure 7. In this structure, one bromine ion is located between two water molecules, and the other is somewhat disordered between three close positions. The outer disordered bromine is the closest to the C3 and C4 carbons, while the interior bromine ion is closer to C1 and C2.

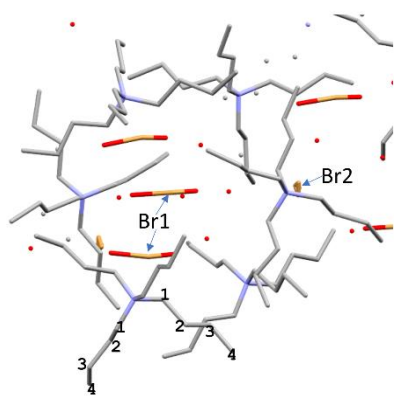

**Figure 7.** The structure of TBAB • 2 1/3H$_2$O showing the arrangement of 6 TBAB molecules around the water and bromine ions. Br2 is slightly disordered. Carbons are labeled on two chains. The distances for this form are in good agreement with the solid-state NMR data.

## 4. Summary and Conclusions

Magic-angle spinning solid-state NMR is an excellent tool to measure highly accurate (down to fractions of an Angstrom) inter-atomic distances. Robust techniques exist to study distances between pairs of nuclei with spins equaling one-half. Recent developments allowed to measure distances also in cases where one of the spins possesses a significantly large quadrupolar coupling constant. This work further extended the NMR distance measurement toolbox. Using split phase-modulated quadrupolar-spin recoupling pulses, combined with a proper hardware modification, we have shown how distances between spins having close Larmor frequencies can be measured, in this particular case between $^{13}$C and $^{81}$Br having a difference of only 7% in their gyromagnetic ratios. Simulations suggest that the method is robust and allows to increase the gap in the phase-modulated recoupling pulse so as to fit extended spin-1/2 pulses if desired, for example in the case of selective pulses. In addition, the recoupling curve under the split-pulse is not sensitive to the strength of the dipolar or quadrupolar coupling strengths, and in all cases can be fit to an analytical Bessel function.

In the experimental example we provide, we show how the distance determination can provide a means to differentiate between two hydrates of tetrabutylammonium bromide by probing distances between the halogen nucleus to all carbons of the butyl chain. It demonstrates the applicability of carbon-bromine distance measurements.

Beyond this particular example, the split-PM RESPDOR experiment can be reliably, accurately and easily used to measure distances between many other spin pairs with close Larmor frequencies where one has a spin larger than one-half, for example $^{13}$C-$^{51}$V, $^{13}$C-$^{27}$Al, $^{13}$C-$^{45}$Sc, $^{117}$Sn-$^{11}$B, and other similar spin pairs that cannot be tuned on a commercial probe, or pulsed simultaneously. Moreover, it is also suitable when a multi-spin system exists since the half-spin can be excited selectively and the existence of several quadrupolar spins can be treated either analytically or by numerical simulations to provide the distances.

## 5. Acknowledgements

This study was funded by the US-Israel binational science foundation, grant #2016169. We thank Dr. Daphna Shimon for help with setting up the WURST-QCPMG static experiments. We thank Prof. Roey Amir for providing us with the TBAB compound.

# SUPPORTING INFORMATION

**Figure S1:** $^{79}$Br WURST-QCPMG static spectra and fits using QUEST software [Perras, Widdifield, Bryce, SSNMR 45-46, 36-44, 2012].

$^{79}$Br WURST-QCPMG spectrum of butyl-triphenylphosphonium bromide (BrBuPPh3). $C_Q$=11.3 MHz, η=0.5. For $^{81}$Br, $C_Q$=9.5 MHz.

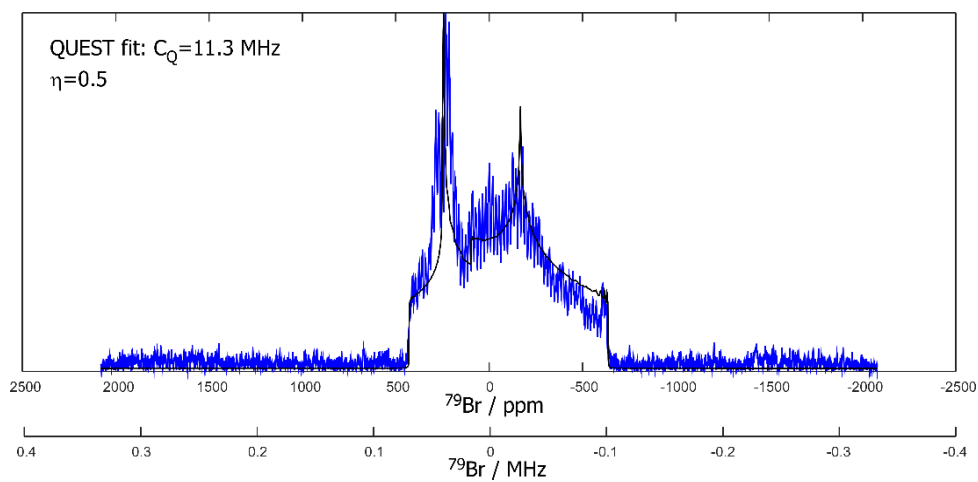

$^{79}$Br WURST-QCPMG spectrum of Tetra-n-butylammonium bromide (TBAB). $C_Q$=12.6 MHz, η=0.16. For $^{81}$Br, $C_Q$=10.5 MHz.

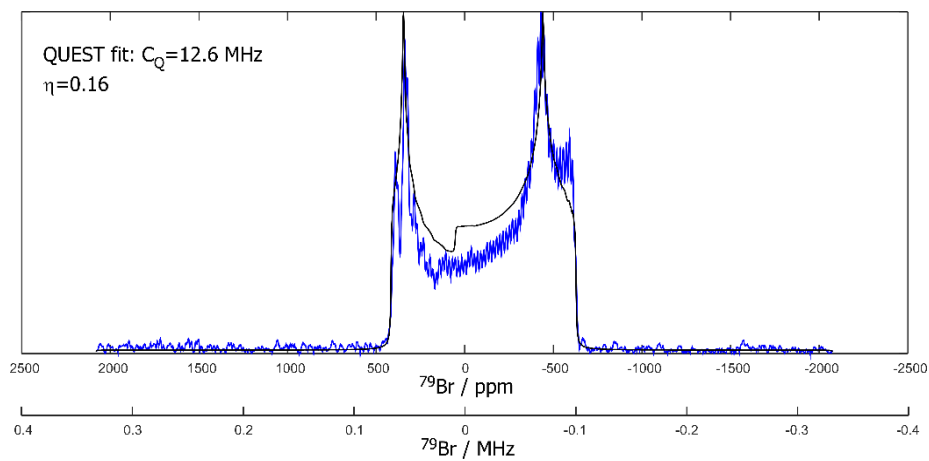

**Figure S2:**

**A.** Split-PM-RESPDOR recoupling curves for a three spin system containing a single $^{13}$C and two Br nuclei at natural abundance – 50% $^{79}$Br-$^{81}$Br pair, 25% $^{79}$Br pair (no $^{81}$Br and therefore no recoupling), 25% $^{81}$Br pair. One $^{13}$C-Br distance is fixed, and the other is at increasing distances (decreasing dipolar coupling).

'x' symbols - simulations of the three-spin system.
'—' Solid black line – simulation of the close bromine (D=130 Hz, ~3.97Å)
'—' Solid blue line – simulation of the remote bromine
'---' Purple dash line – fit of the three-spin system with a single spin-pair.

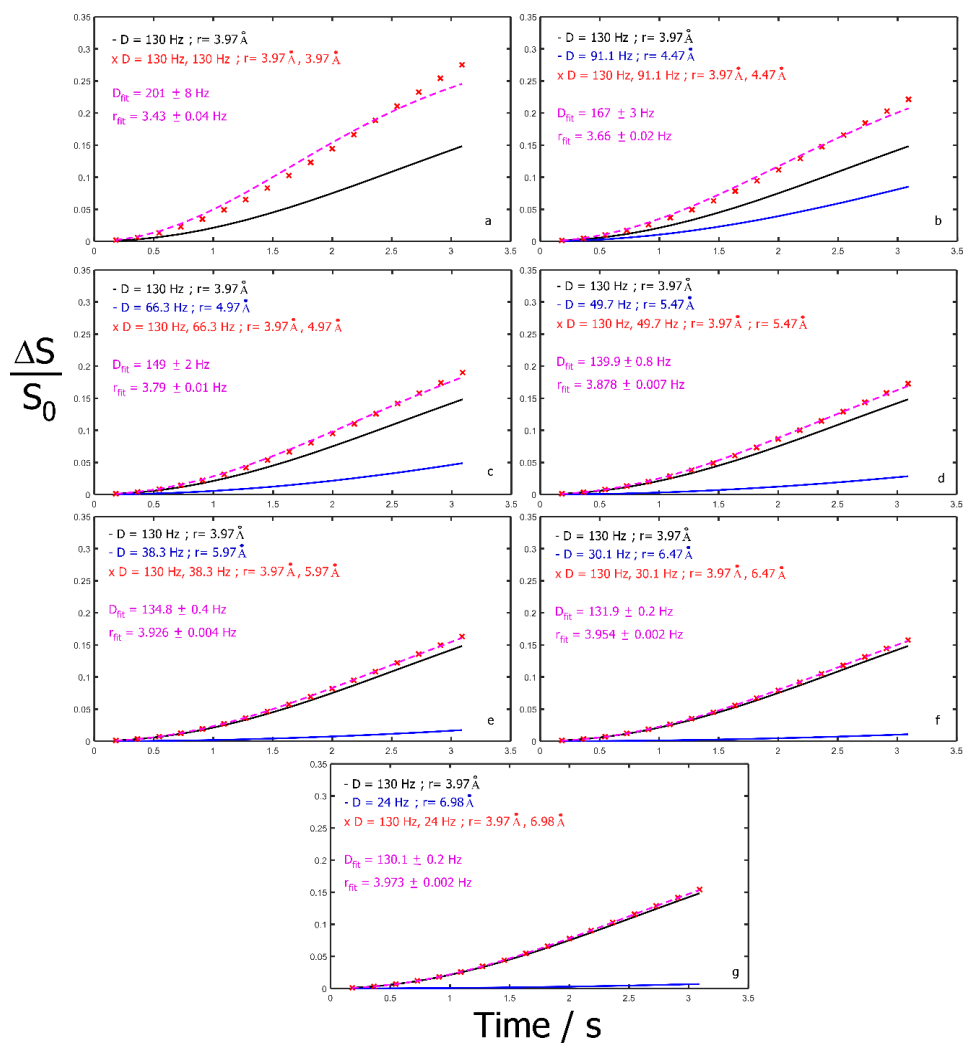

**B.** A plot of $r_{fit}/r$ as a function of difference in distance $\Delta r$, where $r_{fit}$ is the distance extracted by the purple curves above and r=3.97Å. The dash lines indicate $r_{fit}/r$=0.9 and 1.1.

Following similar calculations, data points corresponding to a single bromine with D=121 Hz (r=4.07Å) can be fit with 2Br with D=110.1, 58 Hz (r=4.2Å, 5.2Å).

Similarly, a single Br with D=130 Hz (r=3.97Å) corresponds to 2Br with D=118.3, 61.5 Hz (r=4.1Å, 5.1Å).

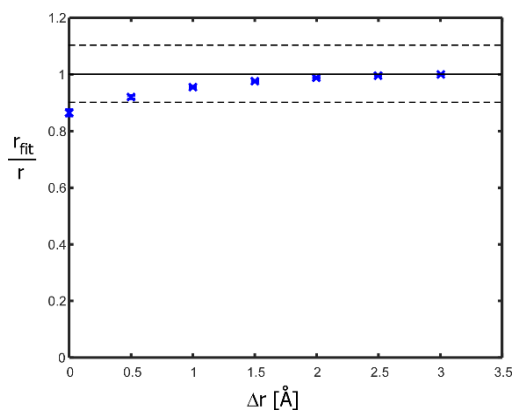

**Figure S3:** Split-PM-RESPDOR recoupling curves for a spin-1/2 coupled to a spin-3/2 having a nuclear quadrupolar coupling constant of 200 MHz. D=130 Hz, the gap is 26.3μs, PM pulse lengths are 10$T_R$ for the red squares, 100$T_R$ for the blue crosses. The solid line is the analytical Bessel function for a spin-3/2 shown in Eq. 1 of the main article.

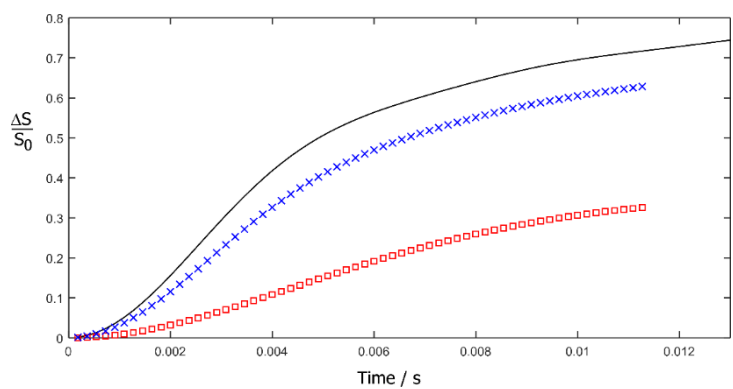

**Figure S4:** Fits of the individual signals of each of C2, C3, C4 following a very slight line broadening of 1 Hz. These signals are shown in the $^{13}$C-CPMAS spectrum shown in Figure 5. Even with this small line broadening, C1 cannot be decomposed to individual peaks. The differences in the fit distance between the different components of each carbon are within 0.1Å. In the main manuscript, the spectrum was line-broadened with 100 Hz and a single distance was obtained for each carbon, as shown in Figure 6. Those fits are indicated here in black.

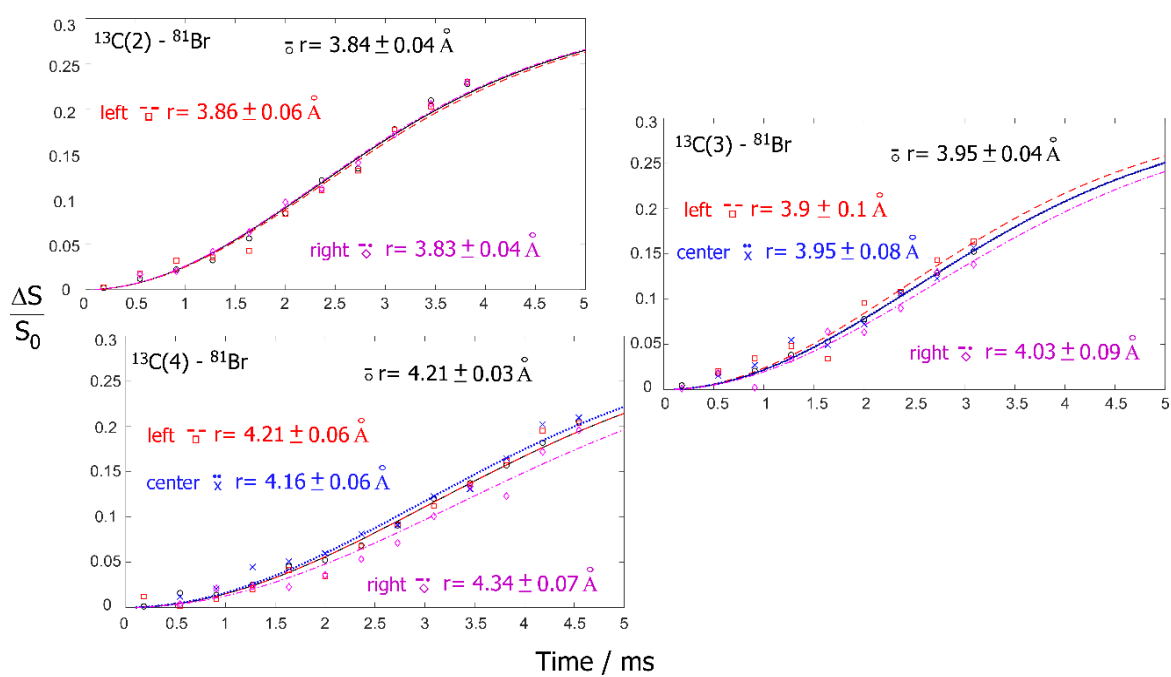

**Item S5**: Bruker pulse program for split-PM-RESPDOR and the corresponding shape files.

Pulse program:

```
;split-PM-RESPDOR pulse program
;based on the papers by Nimerovsky, Makrinich, Goldbourt:
;10.1016/j.jmr.2014.03.003; 10.1016/j.ssnmr.2018.04.001
;Adapted to include a split in the PM pulse
;Avance III version (original II+)
;parameters:
;p3 : proton 90 at power level pl3
;p15 : contact time at pl1 (f1) and pl2 (f2)
;pcpd2 : pulse length in decoupling sequence
;cpdprg2 : cw, tppm (at pl12),
;cpdprg6 : PM phases, first half
;cpdprg7 : PM phases, second half
;pl33: PM pulse power (0W or 120db for S0)
;p31 : pulse length for phase modulated pulses
;p30 : pulse lenth for phase 225 degree
;p32 : gap length
;cnst31 : spin rate >1 kHz
;l0 : 1,3.... has to be odd
;l10 : l0-1 - number of pi pulses in each REDOR block
;l2 : # of rotor periods of coupled spin pulse
;d1 : recycle delay
;d31 : used to check spin rate
;pl1 : X power level for contact
;pl11 : X power level for 180
;sp0 : proton power level during contact
;pl12 : proton power level for decoupling
;pl3 : H90 power level for 90
;spnam0 : file name for variable amplitude CP
;$COMMENT=REDOR experiment, cp for excitation, interleaved acquisition of S and S0 signals, inverse sequence
;$CLASS=Solids
;$DIM=1D
;$TYPE=cross polarisation
;$SUBTYPE=REDOR

;$OWNER=Bruker

;calculate sync. delays
define delay del25
"del25=(0.25s/cnst31)"
define delay del26
"del26=(0.25s/cnst31)-(p12/2)"
define delay del27
"del27=l2*0.5*(1s/cnst31)-(p12)"
define delay del28
"del28=l2*0.5*(1s/cnst31)-(p12/2)-(p32/2)"
define delay del29
"del29=(p32/2)-(p12/2)-(1u)"
;define loopcounter nfid
;"nfid=td1/2"
"d31=1s/cnst31"
;cnst11 : to adjust t=0 for acquisition, if digmod = baseopt
"acqt0=1u*cnst11"
"l10=(l0-1)"
1 ze
  d31
2 10m do:f2
  d1
#include <p15_prot.incl>
                    ;make sure p15 does not exceed 10 msec
                    ;let supervisor change this pulseprogram if
                    ;more is needed
```

```
#include <aq_prot.incl>
                    ;allows max. 50 msec acquisition time, supervisor
                    ;may change  to max. 1s at less than 5 % duty cycle
                    ;and reduced decoupling field
#include <rot_prot.incl>
                    ;protect against misset cnst31, must be >1000
  2u rpp4
  2u rpp9
  2u rpp8
  (p3 pl3 ph1):f2
  (p15 pl1 ph2):f1 (p15:sp0 ph10):f2
  del25  cpds2:f2
if "l0 == 1" goto sk3
3 del26
  (p12 pl11 ph8^):f1
  del26
  lo to 3 times l10
sk3,  del26
  (p12 pl11 ph2):f1
  del26
  del25
 (p12 pl11 ph4^):f1
 1u
 0.1u cpds6:f3
 del28
 0.1u do:f3
 del29
 (p12 pl11 ph4^):f1
 del29
 1u
 0.1u cpds7:f3
 del28
 0.1u do:f3
 del25
if "l0 == 1" goto sk4
4 del26
  (p12 ph9^):f1
  del26
  lo to 4 times l10
sk4,  del26
  (p12 ph2):f1
  del26
  del25
  go=2 ph31
  1m do:f2
  30m wr #0

HaltAcqu, 1m
exit

ph1= 1 3
ph2= 0 0 1 1 2 2 3 3
ph4= 0 1 0 1 1 0 1 0
ph8= 0 1 0 1 1 0 1 0
ph9= 0 1 0 1 1 0 1 0
ph10= 0
ph31= 0 2 1 3 2 0 3 1
```

Shape files:

| Shape before X-channel π pulse (cpds6): | Shape after X-channel π pulse (cpds7): |
|---|---|
| ```
1
p30: 225.00 pl=pl33
p31: 80.51923 pl=pl33
p31: 2.236028 pl=pl33
p31: 244.9407 pl=pl33
p31: 92.52825 pl=pl33
p31: 275.7416 pl=pl33
p31: 98.55024 pl=pl33
p31: 330.2784 pl=pl33
p31: 348.0776 pl=pl33
p31: 126.2695 pl=pl33
p31: 265.2453 pl=pl33
p31: 349.2809 pl=pl33
p31: 23.86148 pl=pl33
p31: 24.11632 pl=pl33
p31: 23.8011 pl=pl33
p31: 92.37856 pl=pl33
p31: 6.6659 pl=pl33
p30: 225.00 pl=pl33
p31: 319.617 pl=pl33
p31: 75.59648 pl=pl33
p31: 242.7034 pl=pl33
p31: 95.8332 pl=pl33
p31: 340.9066 pl=pl33
p31: 226.5105 pl=pl33
p31: 48.00275 pl=pl33
p31: 38.06114 pl=pl33
p31: 5.261435 pl=pl33
p31: 44.36102 pl=pl33
p31: 145.7639 pl=pl33
p31: 303.2499 pl=pl33
p31: 139.6619 pl=pl33
p31: 336.7204 pl=pl33
p31: 95.65318 pl=pl33
p31: 341.3066 pl=pl33
jump to 1
``` | ```
1
p31: 80.51923 pl=pl33
p31: 2.236028 pl=pl33
p31: 244.9407 pl=pl33
p31: 92.52825 pl=pl33
p31: 275.7416 pl=pl33
p31: 98.55024 pl=pl33
p31: 330.2784 pl=pl33
p31: 348.0776 pl=pl33
p31: 126.2695 pl=pl33
p31: 265.2453 pl=pl33
p31: 349.2809 pl=pl33
p31: 23.86148 pl=pl33
p31: 24.11632 pl=pl33
p31: 23.8011 pl=pl33
p31: 92.37856 pl=pl33
p31: 6.6659 pl=pl33
p30: 225.00 pl=pl33
p31: 319.617 pl=pl33
p31: 75.59648 pl=pl33
p31: 242.7034 pl=pl33
p31: 95.8332 pl=pl33
p31: 340.9066 pl=pl33
p31: 226.5105 pl=pl33
p31: 48.00275 pl=pl33
p31: 38.06114 pl=pl33
p31: 5.261435 pl=pl33
p31: 44.36102 pl=pl33
p31: 145.7639 pl=pl33
p31: 303.2499 pl=pl33
p31: 139.6619 pl=pl33
p31: 336.7204 pl=pl33
p31: 95.65318 pl=pl33
p31: 341.3066 pl=pl33
p30: 225.00 pl=pl33
jump to 1
``` |